\documentclass[preprint,showpacs,preprintnumbers,amsmath,amssymb]{revtex4}
\usepackage{graphicx,graphics}
\usepackage{bm}
\usepackage{verbatim}
\usepackage{dcolumn}
\usepackage{epstopdf}

\begin{document}
\title{The propagator for the step potential and delta function potential using the path decomposition expansion}
\author{James M. Yearsley}
\affiliation{Blackett Laboratory\\ Imperial College\\ London SW7 2BZ\\ UK }

\date{15th April 2008}                                           

\begin{abstract} We present a derivation of the propagator for a particle in the presence of the step and delta function potentials. These propagators are known, but we present a direct path integral derivation, based on the path decomposition expansion and the Brownian motion definition of the path integral. The derivation exploits properties of the Catalan numbers, which enumerate certain classes of lattice paths.
\end{abstract}


\maketitle

\section{Introduction}

Calculations involving step and delta function potentials occur in many branches of physics. Step potentials can be used to represent \textquoteleft hard wall' boundary conditions, and are also involved in tunneling calculations. Imaginary step potentials are useful in modeling particle absorbtion, for instance in arrival time problems \cite{muga, halliwell}.  Delta function potentials can be used to model point interactions, especially in the low energy limit where the details of the process are largely independent of the form of the scattering potential \cite{deltapot}. In this paper we present a derivation of the propagators in the vicinity of these potentials by making use of the path decomposition expansion \cite{pdx, halliwell+ortiz, schulman}, and then by appealing to the Brownian motion definition of the path integral. 

\begin{figure}[h]
\includegraphics[scale=0.5]{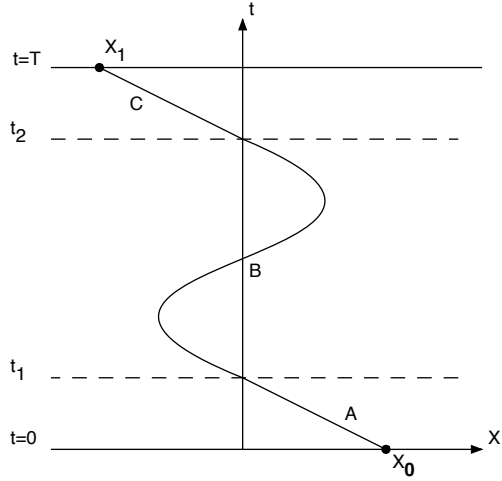}
\caption{{\bf A typical path from $x_{0}$ to $x_{1}$}}\label{pdx}
\end{figure}

We wish to calculate the following propagator

\begin{equation}
g(x_{1},T | x_{0},0)=\int_{x(0)=x_{0}}^{x(T)=x_{1}} \! \mathcal{D}x \; e^{iS}
\end{equation} where
\begin{equation}
S=\int_{0}^{T}dt\left(\frac{m\dot x^{2}}{2} - V(x)\right)
\end{equation}
and the potential $V(x)$ will either be a step potential $V\theta(-x)$, or a delta function potential, $a\,\delta(x)$. (Note that throughout this paper we set $\hbar = 1$.)
To calculate these propagators, the path decomposition expansion \cite{pdx, halliwell+ortiz, schulman} will be employed as follows. A typical path from $x_{0}>0$ to $x_{1}<0$ may cross $x=0$ many times, but the set of paths may be partitioned by their {\it first} or {\it last} crossing times. We therefore split every path into three parts: (A) a restricted part that starts at $x_{0}$ and does not cross $x=0$, but that ends on $x=0$ at time $t_{1}$, (B) an unrestricted part from $x=0$ to $x=0$ that may cross $x=0$ many times and, (C) a further restricted part from $x=0$ to $x_{1}$ that does not re-cross $x=0$, Fig. (\ref{pdx}).
In terms of the first crossing time $t_{1}$,
\begin{equation}
g(x_{1},T|x_{0},0) = \frac{i}{2m} \int_{0}^{T} dt_{1}\; g(x_{1},T|0,t_{1}) \left.\frac{\partial g_{r}}{\partial x}(x,t_{1}|x_{0},0)\right|_{x=0}\label{first}.
\end{equation}
And in terms of the last crossing time $t_{2}$,
\begin{equation}
g(x_{1},T|x_{0},0) = -\frac{i}{2m} \int_{0}^{T} dt_{2}\; \left.\frac{\partial g_{r}}{\partial x}(x_{1},T|x,t_{2})\right|_{x=0} g(0,t_{2}|x_{0},0)\label{last} .
\end{equation}
From the above it is possible to derive a third decomposition, in terms of the first {\it and} last crossing times
\begin{equation}
g(x_{1},T|x_{0},0) = \frac{1}{4m^{2}}\int_{0}^{T}dt_{2}\int_{0}^{t_{2}}dt_{1}   \;  \frac{\partial g_{r}}{\partial x}(x_{1},t|x,t_{2})|_{x=0}    \; g(0,t_{2};0,t_{1})   \; \frac{\partial g_{r}}{\partial x}(x,t_{1};x_{0},0)|_{x=0}\label{path}.
\end{equation}
When $x_{1}$ and $x_{0}$ are both positive, there is an additional contribution from paths which never cross $x=0$ and the path decomposition expansion is, 
\begin{equation}
g(x_{1},T|x_{0},0)=g_{r}(x_{1},T|x_{0},0)-\frac{i}{2m} \int_{0}^{T} dt_{1}\; g(x_{1},T|0,t_{1}) \left.\frac{\partial g_{r}}{\partial x}(x,t_{1}|x_{0},0)\right|_{x=0}\label{hardercase}
\end{equation}

 In the above path decompositions $g_{r}$ are the restricted propagators for intervals (A) and (C), when the particle does not cross the origin.  Since in the regions (A) and (C) the potential is a constant, V (or zero) the restricted propagators in these regions are equal to the restricted propagator for the free particle, weighted by a factor $e^{-V(t_{1}-t_{0})}$. Now the restricted free propagator, $g_{f,r}$, is given by the method of images expression,
\begin{equation}
g_{fr} (x_1, t_{1} |x_0,t_{0}) = \theta (\pm x_1 ) \theta (\pm x_0)
\left( g_f (x_1, t_{1} |x_0,t_{0}) - g_f (-x_1, t_{1} |x_0,t_{0}) \right)\label{images}
\end{equation}
where the $\theta$ functions ensure paths start and end on the same side of $x=0$. So the restricted propagators may therefore be easily computed by weighting the free propagators in Eq. (\ref{images}) according to the region of interest. The problem of calculating the full propagator therefore reduces to that of calculating the partial propagator for the interval where the path crosses the origin $g(0,t_{2}|0,t_{1})$. We will show how these partial propagators may be derived using the Brownian motion definition of the path integral \cite{hartle, gert}. The full propagators may then be obtained with the help of Eqs. (\ref{first}-\ref{path}).

For the step potential the full propagator has been derived in Refs. \cite{car, barut, che, crandall}, but we shall derive the partial propagator, and then direct the reader to Ref. \cite{car} for details of the use of the path decomposition expansion to recover the full propagator.
The partial propagator we will derive is given by
\begin{equation}
g(0,T|0,0)= -i\left(\frac{m}{2 \pi i}\right)^{1/2}\frac{(1-e^{-iVT})}{VT^{3/2}}.
\end{equation}

For the delta function potential we will derive the full propagator and we quote the result from Ref. \cite{delta}
\begin{equation}
g(x_{1},T|x_{0},0)= g_{f}(x_{1},T|x_{0},0)-a\int_{0}^{\infty} \! du\; e^{-amu}g_{f}(|x_{1}|+|x_{0}|+u,T|0,0)
\end{equation}
where
\begin{equation}
g_{f}(x_{1},T|x_{0},0)=\left(\frac{m}{2\pi iT}\right)^{1/2}e^{im(x_{1}-x_{0})^{2}/2T}
\end{equation}
is the free propagator.

\section{The Brownian motion definition of the propagator}
We begin with a review of some of the details of the Brownian motion approach to computing propagators. For more details see Refs. \cite{hartle, gert}. 
The first step is to switch to working with the Euclidean propagator $\overline g$ by means of a Wick rotation and we specialise immediately to the case of $x_{0}=x_{1}=0$.  That is, we wish to calculate
\begin{equation}
\overline g(0,T | 0,0)=\int_{x(0)=0}^{x(T)=0}\! \mathcal{D}x \;e^{-S_{E}}\label{prop}
\end{equation} 
where $S_{E}$ is the Euclidean action given by
\begin{equation}
S_{E}=\int_{0}^{T}dt\left(\frac{m\dot x^{2}}{2} + V(x)\right)\label{prop2}.
\end{equation}
This propagator may be viewed as a conditional probability density for a random walk on the real line. The second step is then to make this integral over paths into a concrete object by defining it as the continuum limit of a discrete sum on a lattice.

\begin{figure}[h]
\begin{center}
\includegraphics[scale=0.5]{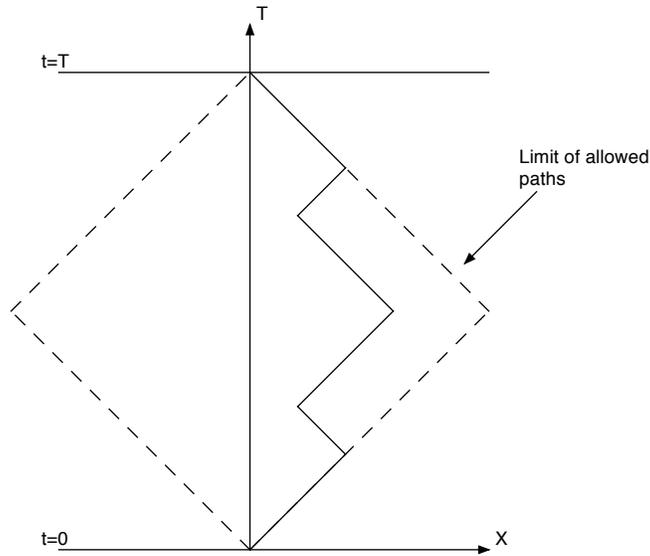}
\caption{{\bf A typical path from $x=0$ to $x-0$. The dashed line shows the area within which all paths must remain.}}\label{typical}
\end{center}
\end{figure}

To establish conventions and demonstrate the basic ideas we compute the case of a free particle, following closely the treatment in Ref. \cite{hartle}. We consider a rectangular lattice with spacing in the time direction of $\epsilon$, and spacing in the $x$ direction of $\eta$, and consider propagation for a time $T=2\epsilon n$, so we have $2n$ steps in our paths (the reason for this choice is that it simplifies a number of later expressions, and avoids clumsy factors of 1/2). The conditional probability $u(0,T|0,0)$ to start at (0,0) and end at (0,T) is given by the number of paths connecting the start and end points, divided by the total number of possible paths. The set of all possible paths is bounded by the extremal paths that take $n$ steps to the left/right, followed by $n$ steps to the right/left, see Fig. (\ref{typical}). Since a path must have the same number of steps to the right as to the left to end up back at $x=0$ we find,
\begin{equation}
u(0,T|0,0) = \frac{1}{2^{2n}}{2n \choose n}
\end{equation}
The Euclidean propagator $\overline g$ is then defined as the continuum limit of $u/2\eta$ where we take $\epsilon,\eta \rightarrow 0$, $n\rightarrow \infty$, keeping $\epsilon/\eta^{2} = m$ and $T=2\epsilon n$ fixed. That is,
\begin{equation}
\overline g_{f}(0,T|0,0):=\lim_{\eta, \epsilon \to 0} (2\eta)^{-1}u(0,T|0,0) = \left(\frac{m}{2\pi T}\right)^{1/2}
\end{equation}
which is the expected result for the Euclidean free propagator.

\section{The step potential}
\begin{figure}[h] 
   \centering
   \includegraphics[width=3in]{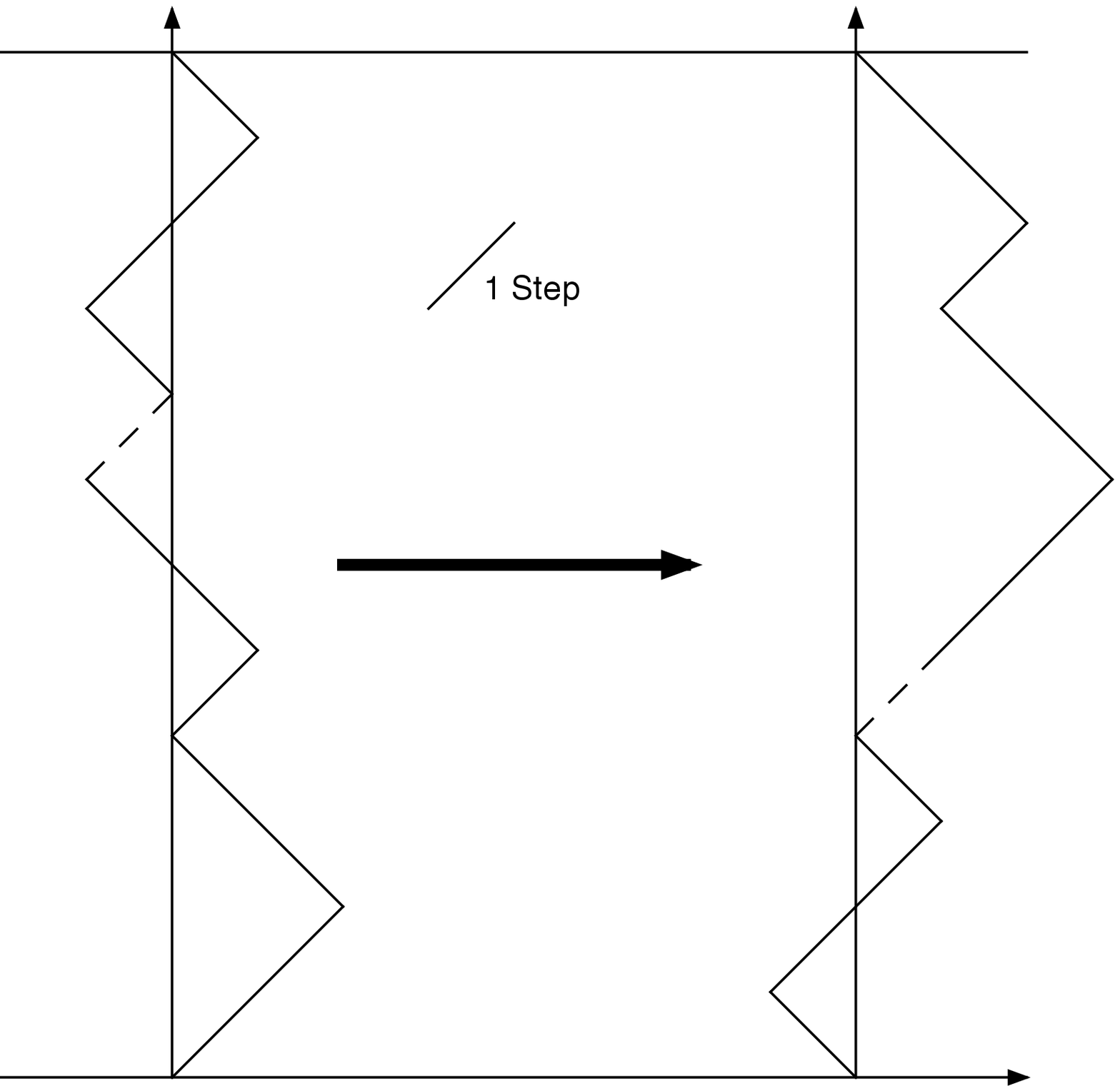} 
   \caption{{\bf Application of the rule to the path on the left produces the path on the right, with 2 timesteps less spent in $x<0$}}\label{pathrule}
\end{figure}

The propagator along the edge of a step potential is given by Eqs. (\ref{prop}) and (\ref{prop2}), with $V(x)=V\theta(-x)$. We can write this as
\begin{equation}
\overline g(0,T|0,0) = \int \mathcal{D}x \exp \left(-\int_{0}^{T}dt \frac{m \dot x^{2}}{2}\right) \exp \left(-V\int_{0}^{T}dt \theta(-x)\right)
\end{equation}
which is similar to the free particle case except that paths are weighted by a factor $e^{-V\tau}$ where
\begin{equation}
\tau = \int_{0}^{T}dt \theta(-x)
\end{equation}
is the length of time spent in $x<0$. In the lattice case the corresponding conditional probability $u(0,T|0,0)$ is given by a sum of paths, each weighted by a similar factor.
This may be written as
\begin{equation}
u(0,T|0,0) = \frac{1}{2^{2n}}\sum_{k=0}^{n} \;n_{k}\; e^{-2k\epsilon V}\label{exp},
\end{equation}
where $n_{k}$ is the number of paths spending a time $2k\epsilon<T$ in the region $x<0$. Expresions for these $n_{k}$ are known and are in fact independent of $k$ \cite{Cat}. They are equal to the Catalan number 
\begin{equation}
C_{n} = \frac{1}{n+1} {2n \choose n}.
\end{equation}
where $2n$ is the total number of time steps. We can see this as follows. First, note that the number of paths that never enter $x<0$ is given by $C_{n}$, this being one definition of the Catalan numbers. Next consider the following mapping on any path spending a time $2\epsilon k<T$ in $x<0$, Fig.(\ref{pathrule}).
\begin{enumerate}
\item Start from t=0, and follow the path until it first crosses $x=0$ (if it doesn't cross then stop, the path is in the set of non-crossing paths.)
\item Follow the path until it comes back to $x=0$ again, note the step at which this happens.
\item Swap the section of path \emph{before} this step with the section \emph{after} it.
\item The new path will now spend 2 fewer time steps in $x<0$.
\end{enumerate}
By repeated application of this mapping, any path can be transformed into one which never crosses $x=0$. The important point about this mapping however, is that it is bijective \cite{Cat}, which proves that the number of paths spending time $2k\epsilon$ in $x<0$ must equal the number of paths that never cross, for any value of $k$. This shows that $n_{k}=C_{n}$ for all $k\leq n$.
So we now have
\begin{eqnarray}
u(0,T|0,0) &=&\frac{1}{2^{2n}}\sum_{k=0}^{n}C_{n}e^{-2k\epsilon V} = \frac{C_{n}}{2^{2n}}\sum_{k=0}^{n}e^{-2k\epsilon V}\label{geo} \\ 
 &=&\frac{C_{n}}{2^{2n}}\left(\frac{1-e^{-2\epsilon(n+1)V}}{1-e^{-2\epsilon V}}\right). 
\end{eqnarray}
This coincides with the free particle case if $V=0$.
Since we plan to take the continuum limit we can Taylor expand the exponential in the denominator to first order in $\epsilon$, and use the following useful assymptotic form for $C_{n}$ \cite{Cat}
\begin{displaymath}
C_{n} \approx \frac{4^{n}}{\sqrt{\pi}n^{3/2}}
\end{displaymath}
to get,
\begin{displaymath}
u(0,T|0,0) \approx  \frac{1}{\sqrt{\pi}2\epsilon n^{3/2}}\left(\frac{1-e^{-2\epsilon(n+1)V}}{V}\right).
\end{displaymath}
Now take the continuum limit as in Section 2 to obtain, after some simple algebra,
\begin{equation}
\overline g(0,T|0,0)= \left(\frac{m}{2 \pi}\right)^{1/2}\frac{(1-e^{-VT})}{VT^{3/2}}
\end{equation}
which implies 
\begin{equation}
g(0,T|0,0) = -i \left(\frac{m}{2 \pi i}\right)^{1/2}\frac{(1-e^{-iVT})}{VT^{3/2}}
\end{equation}
This is our first result, the propagator along the edge of a step potential. The  full propagator from $x_{0}$ to $x_{1}$ may now be obtained by making use of Eq. (\ref{path}) \cite{car}.

\section{The Delta function potential}
We now wish to evaluate Eqs. (\ref{prop}) and (\ref{prop2}) with the  potential given by $V(x) = a\delta(x)$. In a similar way as for the step potential, we can write this as,
\begin{equation}
\overline g(0,T|0,0) = \int \mathcal{D}x \exp \left(-\int_{0}^{T}dt \frac{m \dot x^{2}}{2}\right) \exp \left(-a\int_{0}^{T}dt \delta(x(t))\right)
\end{equation}
which is again similar to the free particle case except that paths are weighted by a factor $e^{-a\sigma}$ where
\begin{equation}
\sigma = \int_{0}^{T}dt \delta(x(t))
\end{equation}
is the number of times a given path crosses $x=0$. We model the delta function as a square potential of width $\eta$, and height $1/\eta$ so that every crossing of $x=0$ is weighted by a factor of $e^{-a\epsilon/\eta}$. Since $\epsilon/\eta^{2}=m$, we can rewrite this as $e^{-\eta am}$ \cite{thanks}.  ($a$ is usually positive, so that this represents suppression, but there is no need to assume this for the derivation.) We take the potential to be located on the left of $x=0$ so that to start and end at $x=0$ involves an even number of crossings. Such choices are only made for convenience and have no significance in the continuum limit. 
The conditional probability density $u$ may be partitioned in a similar way to that for the step potential. However now the partitioning is with respect to the number of times a path crosses the square potential. Since the number of crossings will always be even we partition into classes of paths that cross $2l<2n$ times, so the conditional probability is
\begin{equation}
u(0,T|0,0)=\frac{1}{2^{2n}}\sum_{l=0}^{n}J(n,l)e^{-2lma\eta}\label{summ}
\end{equation}
Where $J(n,l)$ is the number of paths of $2n$ steps that cross the delta potential $2l$ times. It is known that these $J(n,l)$ are given by the $2l^{th}$ convolution of the Catalan numbers \cite{tri}, this can be demonstrated by writing a general path in terms of sums over non-crossing paths.
These convolutions form the diagonal elements in Catalans triangle \cite{tri, tri2, tri3}. We need to know the $n^{th}$ element in the $2l^{th}$ diagonal from the right, which will give us $J(n,l)$.

From the formula for the elements of Catalan's triangle \cite{tri},
\begin{equation}
c(n,k) = \frac{(n+k)!(n-k+1)}{k!((n+1)!}
\end{equation}
it follows that 
\begin{equation}
J(n,l) =c(n+l,n-l)=\frac{(2n)! (2l+1)}{(n-l)!(n+l+1)!}={2n \choose n} \frac{n!n!(2l+1)}{(n-l)!(n+l+1)!}\label{inspect}
\end{equation}
where we have extracted the binomial factor for later convenience.
We cannot perform the summation in Eq.(\ref{summ}) directly as we did for the step function potential, so we take the continuum limit first to leave ourselves with an integral. In order to do this we need a simpler form for the $J$'s. We need the asymptotic form for $n\to \infty$, but plotting Eq. (\ref{inspect}) as a function of $l$ shows that the dominant contribution comes from taking $l\to \infty$ as well. It is possible to derive the following asymptotic form,
\begin{equation}
J(n,l)\sim {2n \choose n}\frac{2l}{n}e^{-l^{2}/n}\label{approx},
\end{equation}
and as this form shows, the maximum value of $J$ occurs when $l$ is of order $\sqrt{n}$. This is a consequence of the Brownian motion origin of the paths, which therefore have Hausdorff dimension 2 \cite{dimension}. The statement that a typical path crosses the origin an infinite number of times is a consequence of this fractal nature of a typical path. 

If we use the asymptotic form for the $J$'s, Eq. (\ref{approx}) and make the change of variable in the partitioning, Eq. (\ref{summ}), $u=2\eta l$ we can turn the summation into an integral,
\begin{equation}
u(0,T|0,0) = \frac{1}{2^{2n}}{2n \choose n} \int_{0}^{2\eta n} \! du \;\frac{u}{2\eta^{2}n}e^{-\frac{u^{2}}{4\eta^{2}n}}e^{-amu}.
\end{equation}
We now take the continuum limit as in Section 2 to obtain
\begin{eqnarray}
\overline g(0,T|0,0) &=&\left(\frac{m^{3}}{2\pi T^{3}}\right)^{1/2}\int_{0}^{\infty}\!du \;u e^{-\frac{mu^{2}}{2T}}e^{-amu}\nonumber \\
& =&\overline g_{f}(0,T|0,0)-a\int_{0}^{\infty}\!du \;e^{-amu}\overline g_{f}(u,T|0,0)\label{5}
\end{eqnarray}
which, after rotating back to real time, yields
\begin{equation}
g(0,T|0,0)= g_{f}(0,T|0,0)-a\int_{0}^{\infty}\!du \;e^{-amu}g_{f}(u,T|0,0).
\end{equation}
We can now use the path decompsition expansion to obtain the propagator for paths which start at $x_{0}$ and end at $x_{1}$. There are 4 cases, depending on the signs of $x_{0}$ and $x_{1}$, we shall present one case here, the others follow in a very similar fashion. 
First note that the free propagator obeys the following,
\begin{eqnarray}
g_{f}(x_{1},T|x_{0},0) &=& \frac{i}{2m} \int_{0}^{T} dt_{1}\; g_{f}(x_{1},T|0,t_{1}) \left.\frac{\partial g_{f,r}}{\partial x}(x,t_{1}|x_{0},0)\right|_{x=0}\label{a}\\
&=& -\frac{i}{2m} \int_{0}^{T} dt_{2}\; \left.\frac{\partial g_{f,r}}{\partial x}(x_{1},T|x,t_{2})\right|_{x=0} g_{f}(0,t_{2}|x_{0},0)\label{b}
\end{eqnarray}
provided $sign(x_{0})=sign(-x_{1})$. ($g_{fr}$ denotes the restricted free propagator.)
(If the start and end points have the same sign then there are paths between the two that never cross $x=0$, so the expression for the path decomposition expansion has to be modified as in Eq.(\ref{hardercase}) \cite{halliwell+ortiz}.)

Secondly note the following identities,
\begin{equation}
g_{f}(a,t_{2}|b,t_{1})=g_{f}(-a,t_{2}|-b,t_{1})=g_{f}(0,t_{2}|b-a,t_{1})=g_{f}(0,t_{2}|a-b,t_{1})\label{point}.
\end{equation} 
which are just expressions of the symmetry of the free propagator.

Consider the case where $x_{0}<0$ and $x_{1}>0$, noting that $u\geq0$  we first use Eq. (\ref{a}) to attach the leg from $(x_{0},0)$,
\begin{eqnarray}
g(0,t_{2}|x_{0},0)&=&g_{f}(0,t_{2}|x_{0},0)-a\int_{0}^{\infty}du e^{-amu} g_{f}(u,t_{2}|x_{0},0)\nonumber\\
&=&g_{f}(0,t_{2}|x_{0},0)-a\int_{0}^{\infty}du e^{-amu} g_{f}(0,t_{2}|x_{0}-u,0)
\end{eqnarray}
where we have used Eq. (\ref{point}) to obtain the second line. Since $x_{0}<0$ we have that $x_{0}-u<0$, so we can attach a leg to $(x_{1},T)$ using Eq. (\ref{b}),
 \begin{eqnarray}
g(x_{1},T|x_{0},0)&=&g_{f}(x_{1},T|x_{0},0)-a\int_{0}^{\infty}du e^{-amu} g_{f}(x_{1},T|x_{0}-u,0)\nonumber\\
&=&g_{f}(x_{1},T|x_{0},0)-a\int_{0}^{\infty}du e^{-amu} g_{f}(|x_{1}|+|x_{0}|+u,T|0,0)\label{hurrah}
\end{eqnarray}
as expected.
Calculation in the other cases proceeds in a very similar manner, and confirms that Eq. (\ref{hurrah}) is valid in all cases.

\section{Acknowledgments}
The author would like to thank J.J. Halliwell for guidance, and for suggesting the problem.


\begin{thebibliography}{99}
\bibitem{muga} J.G. Muga, J.P. Palao, B. Navarro \& I.L. Egusquiza, Phy. Rep. {\bf 395}, 357 (2004)
\bibitem{halliwell} J.J. Halliwell, quant-ph 0801.4308 (2008)
\bibitem{deltapot} S. Albeverio, F. Gesztesy \& R. Hoegh-Krohn, Ann. Inst. H. Poincar\'e {\bf 37}, 1 (1982)
\bibitem{pdx} A. Auerbach \& S. Kivelson, Nucl. Phys. {\bf B257}, 799 (1985)
\bibitem{halliwell+ortiz} J.J. Halliwell \& M.E. Ortiz, Phys. Rev. {\bf D48}, 748 (1993)
\bibitem{schulman} L.S. Schulman, Techniques and Applications of Path Integration (Dover Publications, New York, 2005)
\bibitem{gert} G. Roepstorff, Path Integral Approach to Quantum Physics (Springer-Verlag, Berlin, 1994)
\bibitem{hartle} J.B. Hartle, Phys. Rev. {\bf D37}, 2818 (1988)
\bibitem{barut} A.O.Barut \& I.H. Duru, Phys. Rev. {\bf A38}, 5906 (1988) 
\bibitem{car} T.O. de Carvalho, Phys. Rev. {\bf A47}, 2562 (1993)
\bibitem{che} L. Chetouani, Il Nuovo Cimento, {\bf B108}, 879 (1993)
\bibitem{crandall} R.E. Crandall, J. Phys. {\bf A26}, 3627 (1993)
\bibitem{delta} B. Gaveau \& L.S. Schulman, J. Phys. {\bf A19}, 1833 (1986)
\bibitem{Cat} J.H. van Lint \& R.M. Wilson, A Course In Combinatorics (Cambridge University Press, Cambridge, 1992); R.P. Stanley, Enumerative Combinatorics, Volume 2 (Cambridge University Press, Cambridge, 1999); R.P Stanley, Catalan addendum to Enumerative Combinatorics, Volume 2, available from http://www-math.mit.edu/$\sim$rstan/ec/catadd.pdf
\bibitem{thanks} The author is grateful to J.J. Halliwell for discussions on this point.
\bibitem{tri} E.W. Weisstein, "Catalan's Triangle." From MathWorld--A Wolfram Web Resource. http://mathworld.wolfram.com/CatalansTriangle.html.  
\bibitem{tri2} The On-Line Encyclopedia of Integer Sequences. http://www.research.att.com/$\sim$njas/sequences/A033184. Note that there are several versions of Catalan's Triangle listed in [OEIS], all of which are essentially the same sequence but read by rows or diagonals etc. The simplest version is http://www.research.att.com/$\sim$njas/sequences/A009766 but this entry lacks many of the comments relevant to the current usage.
\bibitem{tri3} L.W. Shapiro, Disc. Math. {\bf 14}, 83 (1976)
\bibitem{dimension} L.F. Abbot \& M.B. Wise, Am. J. Phys. {\bf 49}, 37 (1981)

\end{thebibliography}
\end{document}